\documentclass[10pt,nofootinbib,twocolumn,superscriptaddress,prl]{revtex4-2}
\usepackage{amsmath}
\usepackage{amssymb}
\usepackage{amsfonts}
\usepackage{dsfont}
\usepackage{soul}
\usepackage{hyperref}
\usepackage{color}

\usepackage{tikz,xcolor}
\definecolor{lime}{HTML}{A6CE39}
\DeclareRobustCommand{\orcidicon}{%
	\begin{tikzpicture}
	\draw[lime, fill=lime] (0,0) 
	circle [radius=0.16] 
	node[white] {{\fontfamily{qag}\selectfont \tiny ID}};
	\draw[white, fill=white] (-0.0625,0.095) 
	circle [radius=0.007];
	\end{tikzpicture}
	\hspace{-2mm}
}
\foreach \x in {A, ..., Z}{%
	\expandafter\xdef\csname orcid\x\endcsname{\noexpand\href{https://orcid.org/\csname orcidauthor\x\endcsname}{\noexpand\orcidicon}}
}

\begin{document}
\date{\today}
\title{
Gravity-induced electric currents}

\author{David Edward Bruschi\orcidA{}}
\email{david.edward.bruschi@posteo.net}
\affiliation{Institute for Quantum Computing Analytics (PGI-12), Forschungszentrum J\"ulich, 52425 J\"ulich, Germany}
\affiliation{Theoretical Physics, Universit\"at des Saarlandes, 66123 Saarbr\"ucken, Germany}

\begin{abstract}
We study the generation of an electric current in an ideal conducting coil, immersed in a magnetic field, due to the occurrence of a gravitational perturbation. We show that this effect can be used to detect gravitational waves impinging on the coil as well as gravitational gradients when the coil moves in a static background gravitational field. Our work opens the way to employing induced electric signals to detect dynamical gravitational fields and for gradiometry.
\end{abstract}

\maketitle

\textit{Can gravitational perturbations be harnessed to generate electric currents?}  
In this work we employ electrodynamics in weakly curved spacetime to show that dynamical perturbations of spacetime curvature induce an electric current in a closed conducting wire immersed in a magnetic field. This occurs because the gravitational deformation changes the magnetic flux through the area of the loop. We apply our techniques to the case of impinging gravitational waves or motion of the system in nonuniform gravitational potentials.

The system consists of a closed loop of an ideal conducting material (e.g., a wire) that is immersed in a magnetic field and is placed in a classical curved spacetime with metric $g_{\mu \nu}$.\footnote{We work in (3+1)-dimensions. We use Einstein's summation convention. The metric has signature $(-,+,+,+)$.}  
We assume that all sources of gravitational perturbations are located far from the system and strong local sources of gravity can be treated as systematic terms to be dealt with separately. We can therefore assume that spacetime is weakly curved and gravitational perturbations propagate in flat spacetime within the framework of linearized gravity \cite{Misner:Thorne:1973,Wald:1995,Sathyaprakash:Schutz:2009,Carroll:2019}. The metric reads $g_{\mu \nu}=\eta_{\mu \nu}+\epsilon\,h_{\mu \nu}$, where $\eta_{\mu \nu}$ is the flat Minkowski metric and $\epsilon\ll1$ is a small positive perturbative parameter that controls the magnitude of the effects.\footnote{All quantities of interest in this work have a perturbative expression of the form $A=A^{(0)}+\epsilon\,A^{(1)}+\epsilon^2\,A^{(2)}$.}

The loop forms the boundary $\partial_{\Sigma}$ of a 2-dimensional surface $\Sigma$ that can be thought of as being traversed by magnetic field lines. We specialize to a flat and thin rectangular loop of conducting material that lies in the $xy$-plane and, in the absence of gravitational perturbations, has width $l_{\textrm{x}}$ along the x-axis, height $l_{\textrm{y}}$ along the y-axis, and area $A=l_{\textrm{x}}l_{\textrm{y}}$. This gives the oriented area element $d\Sigma^j=\sqrt{g_2}\,dx\,dy\,\hat{\boldsymbol{z}}$ on the surface $\Sigma$ of the loop. Here $g_2(t,x,y)=1-2\epsilon g^{(1)}_2(t,x,y)+\mathcal{O}(\epsilon^2)$ is the determinant of the metric restricted to the $xy$ plane, and $\hat{\boldsymbol{z}}$ is the unit vector in the $z$ direction.

\begin{figure}[ht!]
\includegraphics[width=0.9\linewidth]{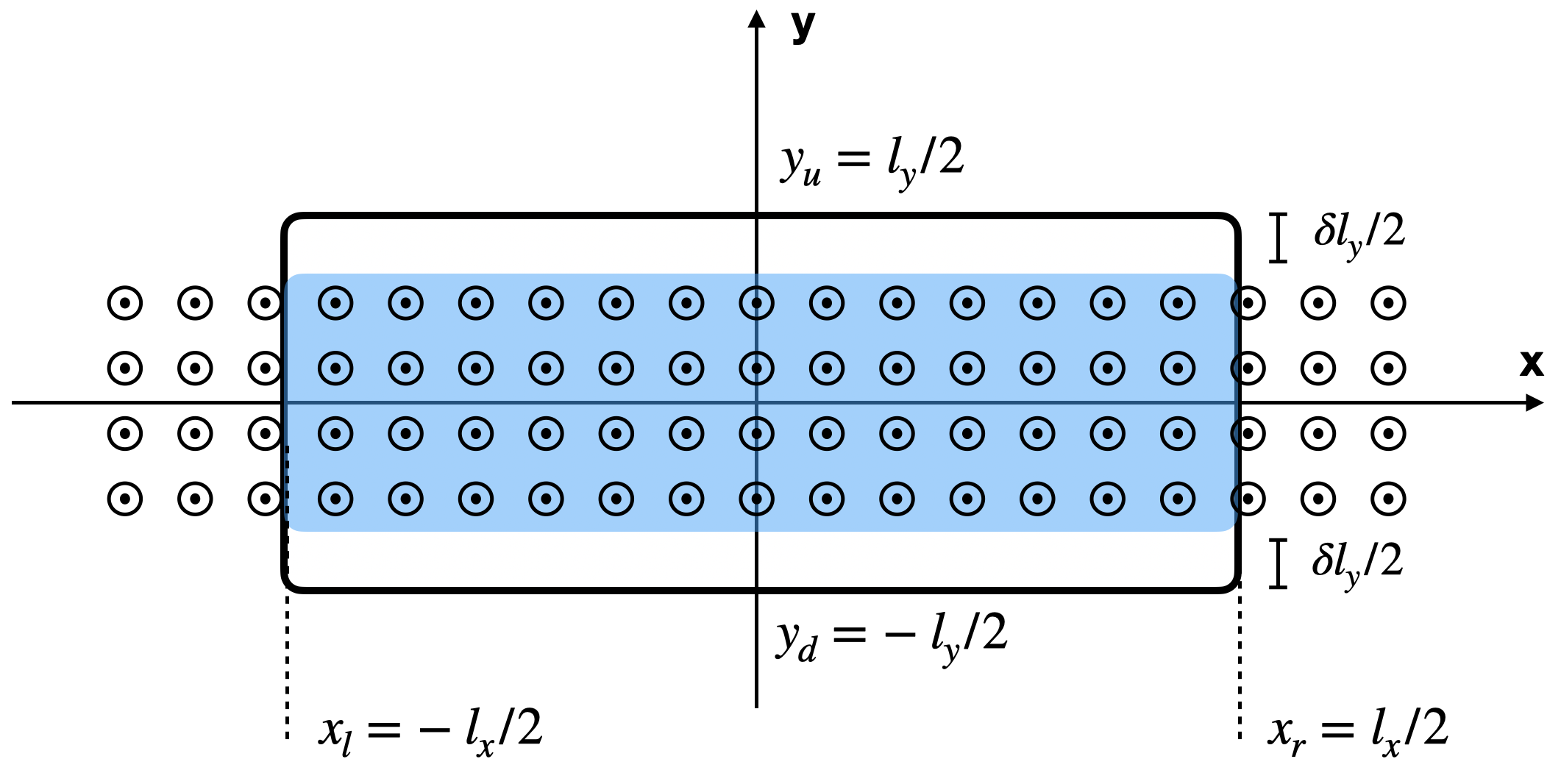}\\
\caption{\textbf{Setup configuration}. The conducting loop is represented by the thick line, and its shape approximates a rectangle of height $l_{\textrm{y}}:=y_{\textrm{d}}-y_{\textrm{u}}$ and width $l_{\textrm{x}}:=x_{\textrm{r}}-x_{\textrm{l}}$. The magnetic field $\boldsymbol{B}=B_0\hat{\mathbf{z}}$ is non vanishing within the light-blue shaded area, which leaves a gap of length $\delta l_{\text{y}}$ on both ends in the $y$-direction between the top and bottom edges of the loop, it is perpendicular to the plane and points outwards.}\label{Figure:Zero}
\end{figure}

Perpendicular to this plane we have a magnetic field $\boldsymbol{B}=B_0 (\vartheta(y+(l_{\textrm{y}}/2-\delta l_{\text{y}}/2))-\vartheta(y-(l_{\textrm{y}}/2-\delta l_{\text{y}}/2)))\,\hat{\boldsymbol{z}}$ pointing in the positive $z$ direction within a strip of length $l_0\gg l_{\textrm{x}}$ in the $x$-direction and height $l_{\textrm{y}}-2\delta l_{\text{y}}$ in the $y$-direction, where $\vartheta(x)$ is the Heaviside step-function. We have therefore assumed that there is a gap in the area occupied initially by the magnetic field in the $y$-direction, while in the $x$-direction the magnetic field extends (far enough) beyond the boundary of the wire. The gap $\delta l_{\text{y}}$ is an external parameter that can be freely adjusted. We will comment on its key role and on the effects of varying its magnitude later in our work. The configuration of the system is depicted in Figure~\ref{Figure:Zero}.

Crucially, the wire is assumed to resist all deformations that arise due to gravitational tidal forces, which is a reasonable assumption for realistic materials in weakly curved spacetime \cite{Lyle:2010}. This translates into fixing the \textit{proper length} of each edge of the wire, defined as the integral $L:=\int_{x^\mu_{\textrm{i}}}^{x^\mu_{\textrm{f}}}\sqrt{-ds^2}$ between two points $x^\mu_{\textrm{i}}$ and $x^\mu_{\textrm{f}}$ evaluated along a trajectory $x^\mu(\lambda)$ parametrized by $\lambda$, where $ds^2=g_{\mu\nu}dx^\mu dx^\nu$. When there is a gravitational perturbation, we can then compute the proper lengths $L_{\textrm{x}}:=\int_{x_{\textrm{l}}}^{x_{\textrm{r}}}\sqrt{-ds^2}|_{y=z=0}$ and $L_{\textrm{y}}:=\int_{y_{\textrm{d}}}^{y_{\textrm{u}}}\sqrt{-ds^2}|_{x=z=0}$ along the edges of the loop aligned with the $x$- and $y$-axis respectively, both at fixed time $t$. We find
\begin{align}\label{proper:lengths:general}
L_{q}(t)=l_{q}\left(1+\epsilon\Lambda_{qq}(t)\right),
\end{align}
where $q=$x,y and we have defined the real time-dependent functions $\Lambda_{\textrm{x}\textrm{x}}(t):=l_{\textrm{x}}^{-1}\int_{x_{\textrm{l}}}^{x_{\textrm{r}}}dx\, h_{\textrm{x}\textrm{x}}(x^\rho)|_{y=z=0}$, and $\Lambda_{\textrm{y}\textrm{y}}(t):=l_{\textrm{y}}^{-1}\int_{y_{\textrm{d}}}^{y_{\textrm{u}}} dy\, h_{\textrm{y}\textrm{y}}(x^\rho)|_{x=z=0}$. In general, the proper lengths $L_{\textrm{x}}(t)$ and $L_{\textrm{y}}(t)$ do not coincide with the coordinate lengths $l_{\textrm{x}}$ and $l_{\textrm{y}}$.\footnote{One instance where this occurs is within regions of (asymptotically) flat spacetime.}
Finally, we define the ``\textit{proper length}'' $L_{\text{loop}}:=2(L_{\textrm{x}}+L_{\textrm{y}})$ and the \textit{proper area} $A:=L_{\textrm{x}}L_{\textrm{y}}$ of the loop, which are constants in time since they are quantities measured physically in the laboratory.

We now wish to compute the time variation of the flux $\Phi_\textrm{M}$ of the magnetic field through the surface $\Sigma$ delimited by the wire, and to relate it to the electric field $j^\Phi_k$ induced on the boundary $\partial_{\Sigma}$ where our wire lies. This can be done by employing electrodynamics in curved spacetime \cite{Thorne:MacDonald:1982,Cabral:Lobo:2017,Cabral:Lobo:2017:v2,Frolov:2020,Brunney:Gradoni:2021} in order to obtain Faraday's law of induction and Lenz's law \cite{Jackson:1999}.  We therefore consider the 4-vector field $A^\nu$ that defines the Faraday tensor $F_{\mu\nu}:=\nabla_\mu A_\nu-\nabla_\nu A_\mu$ whose dynamical equations read $\nabla_\sigma(\nabla^\sigma A_\nu)-\nabla_\sigma(\nabla_\nu A^\sigma)=-R_{\sigma\nu}\,A^\sigma+\mu_0\,J_\nu$, where $J^\nu$ is the conserved four current (i.e., that satisfies $\nabla_\mu J^\mu=0$), and $\nabla_\nu$ is the covariant derivative.
We choose to work in the Lorentz gauge \cite{Wald:1995,Carroll:2019}, which implies that $\nabla_\sigma A^\sigma=0$, and we assume that there are no pre-existing currents in the wire, i.e., $J^\mu=0$. Therefore, our field equation reduces to $\nabla_\sigma(\nabla^\sigma A_\nu)=-R_{\sigma\nu}\,A^\sigma$. A detailed derivation and analysis is left to the Appendix.

To first order in $\epsilon$ we find
\begin{align}\label{main:relations}
\Phi_\textrm{M}:=\int_{\Sigma} d\Sigma^j\, B_j\quad \text{and}\quad \dot\Phi_\textrm{M}=-\oint_{\partial_{\Sigma}} dx^k\, j^\Phi_k,
\end{align}
where the dot stands for derivative with respect to $t$. 

We expect that, due to the rigidity of the wire, the time dependent deformation of spacetime within the area inside the loop will change the flux of the magnetic field ultimately inducing a current. One can think of this process to first order as some field lines entering or exiting the surface as a function of time.
The crucial step, therefore, is to define precisely the surface $\Sigma$ of interest. Since the wire is infinitely rigid, this means that the integration region $\Sigma$ is $|x|\leq L_{\textrm{x}}/2$ and $|y|\leq L_{\textrm{y}}/2$.  

With the area of integration properly defined, we can employ equations \eqref{proper:lengths:general} and \eqref{main:relations} to obtain
\begin{align}\label{main:equation:electromagnetism:three:main}
\left\{
\begin{aligned}
\Phi_\textrm{M}(t)= & (1-\Theta_{\textrm{y}\textrm{y}}(t)\vartheta\left(\Theta_{\textrm{y}\textrm{y}}(t)\right)-\epsilon\Gamma_{\textrm{z}\textrm{z}}(t))\Phi_\textrm{M}(0),\\
\partial_t\Phi_\textrm{M}= &- P\,J_\Phi(t).
\end{aligned} \right.
\end{align}
where we have introduced $\Theta_{\textrm{y}\textrm{y}}(t):=\epsilon_{\textrm{y}}+\epsilon\Lambda_{\textrm{y}\textrm{y}}(t)$, the normalized gap parameter $\epsilon_{\textrm{y}}:=\delta l_{\text{y}}/L_{\textrm{y}}$ as the relative variation in the $y$-direction, the function $\Gamma_{\textrm{z}\textrm{z}}(t):=(1-\epsilon_{\text{y}}\vartheta(\epsilon_{\text{y}})) \int d^2\tilde{k}\,\tilde{g}_2^{(1)\prime}(t,\tilde{k}_{\textrm{x}},\tilde{k}_{\textrm{y}}) \textrm{sinc}(\tilde{k}_{\textrm{x}}) \textrm{sinc}\bigl(\tilde{k}_{\textrm{y}} (1-\epsilon_{\text{y}}\vartheta(\epsilon_{\text{y}}))\bigr)$ for convenience of presentation, and we neglect the $\mathcal{O}(\epsilon^2)$ error that is made in obtaining $\Phi_\textrm{M}(t)$. The quantity $\tilde{g}_2^{(1)\prime}(t,\tilde{k}_{\textrm{x}},\tilde{k}_{\textrm{y}})$ is the Fourier transform of $g_2^{(1)\prime}(t,\textrm{x},\textrm{y})$ in both the $x$ and $y$ dimension, i.e., the determinant of the metric restricted to the surface $z=0$. Furthermore, we have introduced the flux constant $\Phi_\textrm{M}(0)=A B_0$ by approximating $(1-\epsilon_{\textrm{y}})\epsilon\approx\epsilon$ at our working perturbative order. This additional factor can be included if need be.
The expressions \eqref{main:equation:electromagnetism:three:main} give us the explicit expression of the electric field $J_\Phi(t)$ induced in the loop  \cite{Cabral:Lobo:2017:v2}. 

\begin{figure}[ht!]
\includegraphics[width=\linewidth]{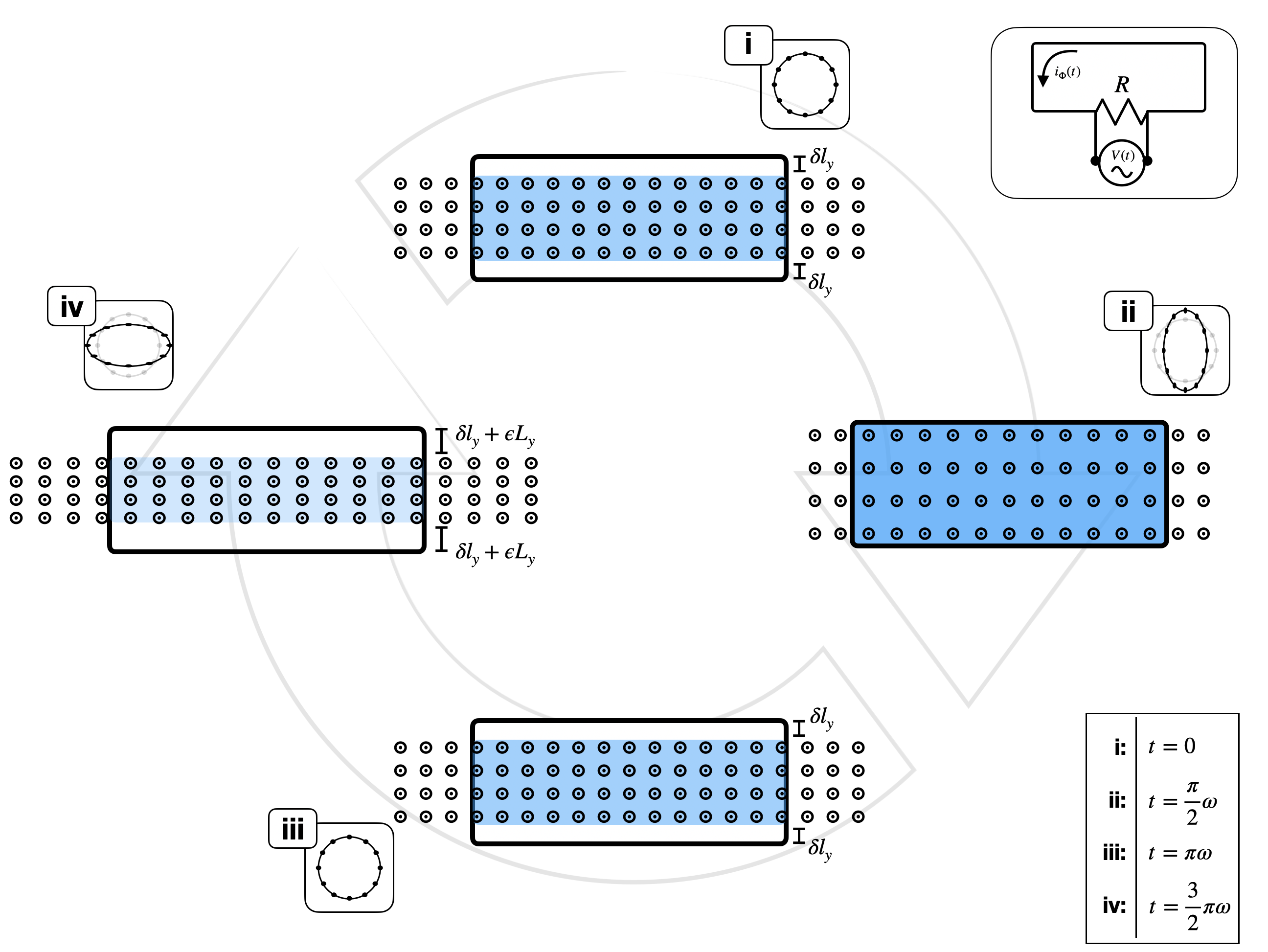}
\caption{\textbf{Effect of an impinging gravitational wave}. The gravitational wave propagates along the $z$-direction and impinges on the system. The effects of spacetime stretching and compression are negligible on the (infinitely) rigid parts of the coil, while they effectively diminish the flux of the magnetic field because of gain or loss of field lines in the $x$-direction.}\label{Figure:One}
\end{figure}

We now proceed to compute the current $i_\Phi(t)$ induced in the wire. To do this we note that the potential $V(\boldsymbol{x}_Q,\boldsymbol{x}_P)$ between to points $\boldsymbol{x}_P$ and $\boldsymbol{x}_Q$ along the path $\boldsymbol{x}(\lambda)$ parametrised by $\lambda$ is given by $V(\boldsymbol{x}_Q,\boldsymbol{x}_P)=-\int_{\boldsymbol{x}_P}^{\boldsymbol{x}_Q}d\lambda\,\dot{\boldsymbol{x}}(\lambda)\cdot\boldsymbol{E}(\boldsymbol{x}(\lambda))$. In our case, we compute the integral along the wire between the two ends of a resistor whose length is very small compared to the length $L_{\text{loop}}=2(L_{\textrm{x}}+L_{\textrm{y}})$, such that its resistance $R$ is approximatively equivalent to that of the loop. In this way, the distance between $\boldsymbol{x}_P$ and $\boldsymbol{x}_Q$ can be approximated by $L_{\text{loop}}$. This choice is not unique but we find it convenient for illustrative purposes. This configuration is depicted in the top right corner in Figure~\ref{Figure:One}. 

In our case the potential reads $V_\Phi(t)=L_{\text{loop}}\,J_\Phi(t)$, and we can compute the current $i_\Phi(t)=L_{\text{loop}} J_\Phi(t)/R$ using the well known law $i=V/R$ that relates current $i$, potential $V$ and resistance $R$, where the latter is the resistance of the wire between the two points considered \cite{Jackson:1999}. In the following we will use the expression $i(t)\equiv i_\Phi(t)$ since we assume that no pre-existing currents are present. We can also use a coil with $N$ spires whose thickness $d$ is much smaller than the characteristic size of variation of the perturbation, i.e., $|\nabla_{\boldsymbol{v}} h(x^\mu)|\ll1/d$ along any unit vector $\boldsymbol{v}$ in the area of interest. In this case, the effective area and the potential between the two electrodes of interest would both increase proportionally to $N$. Furthermore, the resistance would also increase proportionally to $N$. Together, these observations imply that the current $i^{\textrm{coil}}(t)$ along the coil is amplified by the usual factor $N$, see \cite{Jackson:1999}.
All together this gives us
\begin{align}\label{main:result}
i(t)=&N \epsilon\frac{\Phi_\textrm{M}(0)}{R}\left[\dot{\Lambda}_{\textrm{y}\textrm{y}}(t)\,\vartheta\left(\Theta_{\textrm{y}\textrm{y}}(t)\right)+\dot{\Gamma}_{\textrm{z}\textrm{z}}\right],
\end{align}
where we have not included the time derivatives of the Heaviside step-function, which would give Dirac-delta functions. This can be justified by either assuming that $\Theta_{\textrm{y}\textrm{y}}(t)=\epsilon_{\textrm{y}}+\epsilon\Lambda_{\textrm{y}\textrm{y}}(t)\geq0$ for all $t$, or that the magnetic field cannot vanish abruptly in real implementations, and therefore we would need to replace $\vartheta(x)$ with a steep but continuous function (e.g., $\tanh(\alpha x)$ for an appropriate large parameter $\alpha$). Regardless of the approach chosen to tackle this issue, we assume that this boundary-effect can be ignored for all purposes of this work. 
The expression \eqref{main:result} for the induced current is our main result.

We now look for the peak achievable current $i^{\textrm{coil}}_{\textrm{max}}$. We start by optimizing with respect to time, which requires us to seek the stationary points of the quantity $Q(t):=\dot{\Lambda}_{\textrm{y}\textrm{y}}(t)\,\Theta_{\textrm{y}\textrm{y}}(t)+\dot{\Gamma}_{\textrm{z}\textrm{z}}=\dot{\Lambda}_{\textrm{y}\textrm{y}}(t)\,\vartheta\left(\epsilon_{\textrm{y}}+\epsilon\Lambda_{\textrm{y}\textrm{y}}(t)\right)+\dot{\Gamma}_{\textrm{z}\textrm{z}}$. This is in principle easy to do by solving the equation $0=\dot{Q}(t)=\ddot{\Lambda}_{\textrm{y}\textrm{y}}(t)\,\vartheta\left(\epsilon_{\textrm{y}}+\epsilon\Lambda_{\textrm{y}\textrm{y}}(t)\right)+\ddot{\Gamma}_{\textrm{z}\textrm{z}}$ with respect to time, and ignoring once more the problem of the derivatives of $\vartheta(x)$. We obtain the time(s) $t_{\textrm{max}}$ at which the maximum occurs, and consequently $Q(t_{\textrm{max}})$.  We can then optimize our current further by fixing the total length $L_{\text{loop}}=2(L_{\textrm{x}}+L_{\textrm{y}})$ of the single loop in the coil and varying the lengths of the edges. In our case, we can keep $L_{\text{loop}}$ constant, express $L_{\textrm{y}}$ in terms of $L_{\textrm{x}}$ and then maximize the expression \eqref{main:result} as a function of $L_{\textrm{x}}$. Here, we assume that the resistance $R$ is a function of $L_{\text{loop}}$ solely and is therefore constant. It is immediate to see that the maximum of the quantity $A=L_{\textrm{x}}L_{\textrm{y}}$ occurs for the square configuration where $L_{\textrm{x}}=L_{\textrm{y}}=L_{\text{loop}}/4$. Putting all together, we finally obtain
 \begin{align}\label{main:result:peak}
i^{\textrm{coil}}_{\textrm{max}}=&\frac{1}{16}N\ \frac{B_0 L_{\text{loop}}^2}{R} Q(t_{\textrm{max}})\,\epsilon.
\end{align}
The only remaining nontrivial dependence of $i^{\textrm{coil}}_{\textrm{max}}$ is on the shape of the gravitational perturbation encoded in $Q(t_{\textrm{max}})$, which can be obtained analytically or numerically depending on the specific case of interest.

We proceed by presenting three different applications of the results obtained above:
\\
\textit{I. Radiative gravitational perturbations}---Here we focus on dynamical radiative gravitational perturbations that propagate through spacetime. 
Gauge freedom can be used to show that the perturbation $h_{\mu \nu}$ propagates as a superposition of plane waves $e^{k_\sigma x^\sigma}$ far from the source \cite{Misner:Thorne:1973,Carroll:2019}. We assume that we work in the standard transverse-traceless (TT) gauge, which gives us
\begin{align} \label{metric:perturbation}
h_{\mu \nu}(x^\sigma)=\Re\int d^3k\,C_{\mu\nu}(\bold{k}) e^{k_\sigma x^\sigma},
\end{align}
where the four-momentum vector reads $k^\mu=(k_0,\bold{k})$ and satisfies $k_\mu k^\mu=-k_0^2+|\bold{k}|^2=0$, the coefficients $C_{\mu\nu}(\bold{k})$ depend on the spatial momentum $\bold{k}$ and satisfy $C_{00}=C_{0j}=k^\mu C_{\mu\nu}(\bold{k})=0$ and $C:=$Tr$(C_{ij})=C^i{}_i=0$.

We can simplify our equations since $g_{00}=-1$ and $g_{0k}=0$. This implies that we can foliate spacetime using using the timelike vector $\partial_t$, and it is easy to show that det$(g_3)=1+2\Re\int d^3k\,C^j{}_j(\bold{k}) e^{k_\sigma x^\sigma}=1+\mathcal{O}(\epsilon^2)$, where $g_3$ is the determinant of the spatial part of the metric. 
We then compute $g_2(t,x,y)$ for our choice of surface orientation and find $g_2(t,x,y)=1-2\epsilon\Re\int d^3k\,C_{\textrm{z}\textrm{z}}(\bold{k})\,e^{k_\sigma x^\sigma}|_{z=0}+\mathcal{O}(\epsilon^2)$. 
We conclude that $d\Sigma^j$ can have first order contributions that depend on the component of the metric perturbation perpendicular to $\Sigma$. 

The proper lengths \eqref{proper:lengths:general}, using $q=$x,y, for this case read
{\small
\begin{align}\label{proper:lengths:dynamical}
L_{q}(t)=&l_{q}\left(1+\epsilon\Re\int d^3k \, C_{qq}(\bold{k})e^{-i|\bold{k}|t}\textrm{sinc}\left(\frac{k_{\textrm{q}}l_{\textrm{q}}}{2}\right)\right).
\end{align}
}

We repeat the computations done above and obtain the same formal expressions \eqref{main:equation:electromagnetism:three:main}, where in this case we need the function $\Gamma_{\textrm{z}\textrm{z}}(t):=(1-\epsilon_{\text{y}}\vartheta(\epsilon_{\text{y}}))\,\Re\int d^3\tilde{k}\,\tilde{C}'_{\textrm{z}\textrm{z}}(\tilde{\bold{k}})e^{-i\tilde{\omega}_{\tilde{\bold{k}}}t}\textrm{sinc}\bigl(\tilde{k}_{\textrm{x}}\bigr)\text{sinc}\bigl((1-\epsilon_{\text{y}}\vartheta(\epsilon_{\text{y}}))\tilde{k}_\text{y}\bigr)$. Again, we neglect $\mathcal{O}(\epsilon^2)$ and $\mathcal{O}(\epsilon_\text{y}\epsilon)$ errors that are made in obtaining $\Phi_\textrm{M}(t)$, and use \eqref{main:equation:electromagnetism:three:main} to obtain the explicit expression for $J_\Phi(t)$. 
\\
\textit{II. Gravitational waves}---We next study the case of an impinging gravitational wave \cite{Einstein:1918,Misner:Thorne:1973}, which is of great physical importance and current interest in particular since the recent successful detections of gravitational waves by the LIGO collaboration \cite{Abbott:Abbott:2016,Cahillane:Mansell:2022}.

We follow standard derivations for the metric perturbation of a gravitational wave \cite{Misner:Thorne:1973,Carroll:2019}. In the TT gauge, the radiative perturbation \eqref{metric:perturbation} reduces to one where $C_{\mu3}=0$, the wave can be assumed to have sharp momentum and the only nonzero components are therefore $h_{\textrm{xx}}=-h_{\textrm{yy}}$ and $h_{\textrm{x}\textrm{y}}$. The diagonal and off-diagonal components are independent and represent two different sets of oscillation degrees of freedom, i.e., two different polarizations of the wave \cite{Misner:Thorne:1973,Carroll:2019}. Here we assume, for the sake of simplicity and without loss of generality, that $h_{\textrm{x}\textrm{y}}=0$ and, therefore, we have $h_{\textrm{xx}}=-h_{\textrm{yy}}=h_+(t)=\sin(\omega_0  (z-ct))$.
The two proper lengths \eqref{proper:lengths:dynamical} reduce to $L_{\textrm{x}}=(1+\epsilon h_+(t))l_{\textrm{x}}$ and $L_{\textrm{y}}=(1-\epsilon h_+(t))l_{\textrm{y}}$. Notice that the (proper) area $A$ satisfies $A=l_{\textrm{x}}l_{\textrm{y}}+\mathcal{O}(\epsilon^2)$ as expected for this case, since $C_{\textrm{z}\textrm{z}}=0$ and therefore $g_2(t,x,y)=\mathcal{O}(\epsilon^2)$. The (proper) length $L_{\text{loop}}$ of the each loop reads $L_{\text{loop}}=2(l_{\textrm{x}}+l_{\textrm{y}}+\epsilon h_+(t)(l_{\textrm{x}}-l_{\textrm{y}}))$ to first order.

Since we have $h_+(t)=\sin(\omega_0  (z/c-t))$, it is natural to fix $\delta l_{\text{y}}$ via $\delta l_{\text{y}}/L_{\textrm{y}}\equiv\epsilon_{\textrm{y}}=\epsilon$. This is equivalent to saying that the vertical gap left in the area occupied by the magnetic field is exactly equal to the maximum variation of the area itself in that direction. A depiction of the system as a function of time can be found in Figure~\ref{Figure:One}.

The magnetic flux reduces to $\Phi_\textrm{M}(t)=\bigl[1+\epsilon h_+(t)\,\vartheta\left(\epsilon-\epsilon h_+(t)\right)\bigr]\Phi_\textrm{M}(0)=(1+h_+(t)\epsilon)\Phi_\textrm{M}(0)$, which immediately gives us the current and the magnitude of the optimized peak current respectively:
\begin{align}\label{main:result:almost}
i(t)=-\dot{h}_+(t)\,\frac{\Phi_\textrm{M}(0)}{R}\,\epsilon,\,\quad\, i^{\textrm{coil}}_{\textrm{max}}=\,\frac{N}{16}\frac{\omega_0 B_0 L_{\textrm{loop}}^2}{R}\,\epsilon.
\end{align}
\\
\textit{III. Motion of the system}---In the last example, we change the perspective and assume that the background gravitational field is static. This can be a good approximation for setups where measurements are done very fast compared to the time-scales of the background field dynamics. The prototypical example is the measurement of the gravitational field gradient at a fixed location. 

\begin{figure}[ht!]
\includegraphics[width=0.8\linewidth]{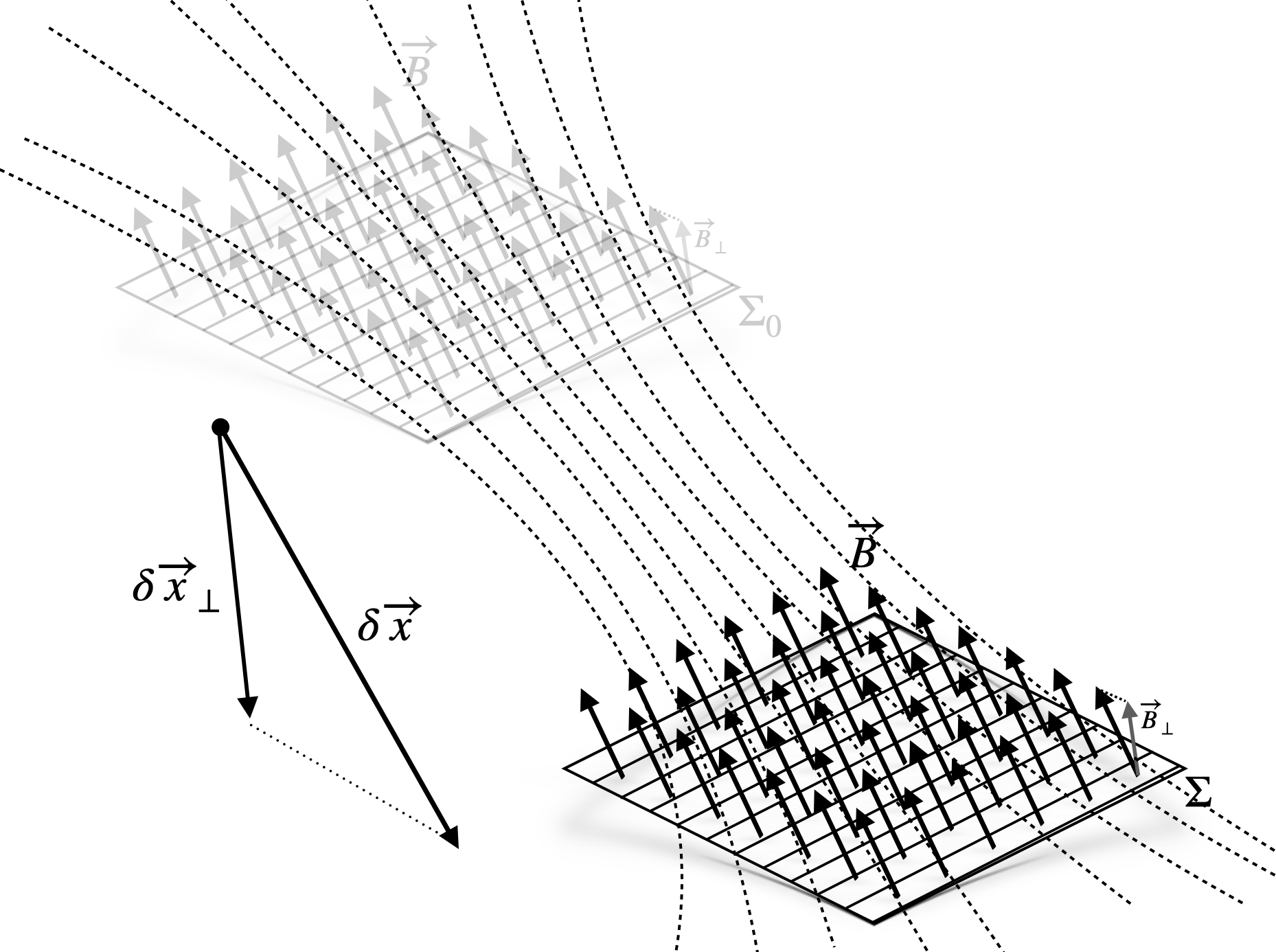}
\caption{\textbf{Pictorial representation of the moving wire}. The wire moves by a small distance $\delta\overset{\rightarrow}{x}\equiv\delta\boldsymbol{x}=\boldsymbol{v}dt$ in a time $dt$, while the gravitational potential is static and position-dependent. Dotted lines indicate gravitational iso-potential.}\label{Figure:Five}
\end{figure}

The flat rectangular loop lies on the initial surface $\Sigma_0$ and is static for $t<0$. It begins to move in a chosen direction at $t=0$. It follows from \eqref{main:equation:electromagnetism:three:main} that the flux $\Phi_\textrm{M}(t)$ is constant for $t\leq0$ and therefore both the induced electric field $J_\Phi(t)$ and induced current $i(t)$ vanish. Nevertheless, once the loop begins moving, the relation between the coordinate lengths and the proper lengths of the edges of the loop, as well as the determinant $g^{\Sigma}_2$ evaluated on the surface $\Sigma$, become time-dependent. This occurs because they are obtained at the particular hypersurface that defines the location of the loop. If the latter changes position, inevitably these quantities need to be calculated ex novo at each instant of time $t$. This does not occur, for example, when the field is homogeneous and the loop travels at a constant velocity without rotations. A generic illustration of the process is found in Figure~\ref{Figure:Five}.

Changes of the determinant $g^{\Sigma}_2(t)$ due to an infinitesimal deviation $\delta \boldsymbol{x}=\dot{\boldsymbol{x}}dt=\boldsymbol{v}dt$ in an arbitrary direction from $\Sigma_0$ at $t_0$ to $\Sigma_{t}$ at $t=t_0+dt$ are implemented by 
\begin{align}\label{square:root:g2:example:3}
g^{\Sigma_{t}}_2(t)=\left(1+\left.\nabla \ln(g_2^{\Sigma_0}(t_0))\right|_{\Sigma_0}\cdot\boldsymbol{v}\,dt\right)g_2^{\Sigma_0}(t_0),
\end{align}
where $\nabla$ is the gradient. One can then chose a particular time dependence $\delta \boldsymbol{x}(t)$ for the small deviation and integrate all infinitesimal expressions to obtain the functions $\Lambda_{11}(t)$, $\Lambda_{22}(t)$ and $\Gamma_{\textrm{z}\textrm{z}}(t)$ at an arbitrary distance in the vicinity of the initial surface $\Sigma_0$.
This, in turn, affects the expressions of proper length and therefore the explicit expression of the current. Numerical integration might be necessary to obtain $g^{\Sigma}_2$, $\Lambda_{11}(t)$ and $\Lambda_{22}(t)$ analytically. 
It is clear that an effect occurs when $\nabla \ln(g_2^{\Sigma})|_{\Sigma_0}\cdot\boldsymbol{v}\neq0$. Examples of motion are: (a) $\delta \boldsymbol{x}(t)=\boldsymbol{v} t$ with velocity parameter $\boldsymbol{v}$; (b) $\delta \boldsymbol{x}(t)=\frac{1}{2}\boldsymbol{a} t^2$ with acceleration parameter $\boldsymbol{a}$; and (c) $\delta \boldsymbol{x}(t)=\Delta \boldsymbol{x}\,\cos(\omega t)$ with oscillation amplitude $\Delta \boldsymbol{x}$ and frequency $\omega$.

As a simple application of this scenario we can assume that the metric is (effectively) constant in the $xy$ plane and that the loop lies on the surface $\Sigma_0$ defined by $z=0$. It follows that the proper length and coordinate length of the edges coincide and therefore the only dependence on time in \eqref{main:result} comes through the term $\Gamma_{\textrm{z}\textrm{z}}$. Thus, we obtain $i(t)=N\frac{\Phi_\textrm{M}(0)}{R}\,\dot{\Gamma}_{\textrm{z}\textrm{z}}\,\epsilon$.

We now evaluate the magnitude of the effects and focus on applications for detection of impinging gravitational waves \cite{Abbott:Abbott:2020,Cahillane:Mansell:2022}. We assume that the wire has section $S=10^{-5}\textrm{m}^2$ and length $10^{-2}\textrm{m}$, is made of silver, and the magnetic field is $B_0=1$ T. This implies that the resistivity is $\rho=1.6\times10^{-8}\frac{\textrm{s}^3 \textrm{A}^2}{\textrm{kg}\, \textrm{m}^3}$ at room temperature and therefore the magnitude of the peak current is $i^{\textrm{coil}}_{\textrm{max}}\approx4\, N\,\omega_0\,\epsilon$ (A$\cdot$s).
In general, gravitational waves will have very small amplitudes, which is the main factor of their elusiveness \cite{Moore:Cole:2014}. Although the parameter $\epsilon$ depends on the frequency $\omega_0$ itself, we can expect it to take values of $\epsilon\sim10^{-20}$ or less. This gives a crude upper bound $i^{\textrm{coil}}_{\textrm{max}}<4\times10^{-20} N\,\omega_0$ (A$\cdot$s).

A few comments on our scheme and results are in place. We note that the existence of a first order effect is directly related to the gap of size $\delta l_{\text{y}}$ when $\dot{\Gamma}_{\textrm{z}\textrm{z}}=0$. The full expression for $\Phi_\textrm{M}(t)$ in \eqref{main:equation:electromagnetism:three:main} shows that, when $\epsilon_{\textrm{y}}\leq-\epsilon$, we always have $\Phi_\textrm{M}(t)=0$. This can be intuitively understood by the fact that the loss of flux in one direction is compensated by the gain of flux in the other to $\mathcal{O}(\epsilon)$. Therefore, we expect only a second order effect, i.e., an area effect. On the other hand, when $\epsilon_{\textrm{y}}>\epsilon\Lambda^\text{max}_{\textrm{y}\textrm{y}}$, where $\Lambda^\text{max}_{\textrm{y}\textrm{y}}$ is the largest value of $\Lambda_{\textrm{y}\textrm{y}}(t)$, there is no additional gain in the magnitude of the induced current with respect to the case where $\epsilon_{\textrm{y}}=\epsilon\Lambda^\text{max}_{\textrm{y}\textrm{y}}$. It is also not difficult to show that, when $\epsilon_{\textrm{y}}<\epsilon\Lambda^\text{max}_{\textrm{y}\textrm{y}}$, the first contribution to the current \eqref{main:result}, as well as the overall current in the case of gravitational waves, vanishes for times $t$ for which $\Lambda_{\textrm{y}\textrm{y}}(t)<-\epsilon_{\text{y}}/\epsilon$. This also shows us that, in the case of gravitational waves, when $\delta l_{\text{y}}=0$ there is an effect at first order only for one half of the period (e.g., stages (iii) and (iv) of Figure~\ref{Figure:One}). 
We also note that we have ignored the problem of the generation of the external magnetic field. The field need not be homogenous--which is an assumption that we have made for the sake of simplicity--but can be engineered to improve the magnitude of the effects. A case by case optimization is required. Regardless of the design, one or more sources of the field must be present and we expect them to be close to the surface of interest. We also have assumed implicitly, in the third example, that the system moves rigidly together with the source of the magnetic field.
It remains to be seen if dynamical perturbations of the gravitational field or motion would affect the physical processes that generate the magnetic field to a degree that can compensate and therefore reduce the induced current. We can say that if any such effect is the result of a second-order process, we can safely ignore it. Regardless, further work on electrodynamics within media in curved spacetime is required to address this issue \cite{Pustovoit:Gladyshev:2019}.
Furthermore, we note that, in order to amplify the peak current $i^{\textrm{coil}}_{\textrm{max}}$, we have to use materials with lower resistivity, and increase the background magnetic field, the number of spires, the section of the wire and, of course, the strength of the perturbation. In the case gravitational waves, high frequencies are desirable.
Finally, we emphasize that our work is related to many studies and developments of this line of research. These include electrodynamics in curved spacetime \cite{Cooperstock:1968,Williams:2012,Cabral:Lobo:2017:v2,Jones:Singleton:2019,Keller:Hively:2019,Lindgren:Liukkonen:2021,Obukhov:2021,Hadj:Dolan:2022,Bunney:Gradoni:2022,Ruggiero:Ortolan:2023}, gravitational-wave interactions with electromagnetic fields and sources \cite{Gertsenshtein:1962,Boccaletti:DeSabbata:1970,Reinhardt:1971,Halpern:1972,Zeldovich:1974,Fortini:Gualdi:1991,Cabral:Lobo:2017}, understanding curved spacetime electrodynamics in media \cite{Gonano:Zich:2015,Cabral:Lobo:2017:v2,Pustovoit:Gladyshev:2019}, applications for gravitational-wave detection \cite{Inan:Thompson:2017} in particular with superconducting circuits \cite{Cooperstock:1968,Schrader:1984,Fortini:Montanari:1996,Chiao:2007,Gulian:Foreman:2021}, and gradiometry \cite{Papini:1967,Snadden:McGuirk:1998,Carraz:Siemes:2014,Carraz:Siemes:2014,Evstifeev:2017,Griggs:Moody:2017,Veryaskin:2018,Stray:Lamb:2022,Bidel:Zahzam:2022} to name a few. We hope that our work will stimulate additional research at the intersection of some of these avenues.

To conclude, we have shown that weak dynamical gravitational perturbations can induce a current in a closed loop of conducting material. As a complementary result, motion of the loop in a weak static gravitational field has the same effect. The magnitude of the current is linear or quadratic in the strength of the perturbation depending on the specific design of the system. The crucial aspect in this dramatic difference is the fact that the background magnetic field might not permeate completely the area occupied by the loop in one spatial direction thereby leaving a small gap of judiciously chosen size. Optimization of the overall effect with respect to the free parameters of the problem, as well as the development of better geometric designs, remains an open task. We believe that our work opens the way to gradiometry and sensing of dynamical gravitational fields via induced electric signals.


\textit{Acknowledgements}---We thank Carlo Andrea Gonano, Frank K. Wilhelm, Leila Khouri, Valente Pranubon, Jorma Louko and Luis Andr\'an Alan\'is Rodr\'iguez for useful comments. We extend special thanks to Andreas Wolfgang Schell for considerations on avenues for concrete implementation of this work. D.E.B. acknowledges support from the joint project No. 13N15685 ``German Quantum Computer based on Superconducting Qubits (GeQCoS)'' sponsored by the German Federal Ministry of Education and Research (BMBF) within the funding framework “Quantum technologies–from basic research to the market”.


\bibliographystyle{apsrev4-2}
\bibliography{GraviCurrentBib}


\appendix

\onecolumngrid

\newpage

\section{SETUP}

Here we derive our main expression. In the absence of the gravitational wave our rectangle coil has width $l_{\textrm{x}}$, height $l_{\textrm{y}}$ and area $A:=l_{\textrm{x}}l_{\textrm{y}}$, and it lies in the $z=0$ plane. This assumption can be lifted to a more general situation. We specialize here for the sake of clarity and simplicity. There is also a constant magnetic field $\boldsymbol{B}=B_0 (\vartheta(y+(l_{\textrm{y}}/2-\delta l_{\text{y}}/2))-\vartheta(y-(l_{\textrm{y}}/2-\delta l_{\text{y}}/2)))\,\hat{\boldsymbol{z}}$ pointing in the positive $z$ direction within a strip of length $l_0\gg l_{\textrm{x}}$ and height $l_{\textrm{y}}-2\delta l_{\text{y}}$, where $\vartheta(x)$ is the Heaviside theta function defined by $\theta(x)=0$ for $x<0$ and $\vartheta(x)=1$ for $x\geq0$. 

\begin{figure}[ht!]
\includegraphics[width=0.6\linewidth]{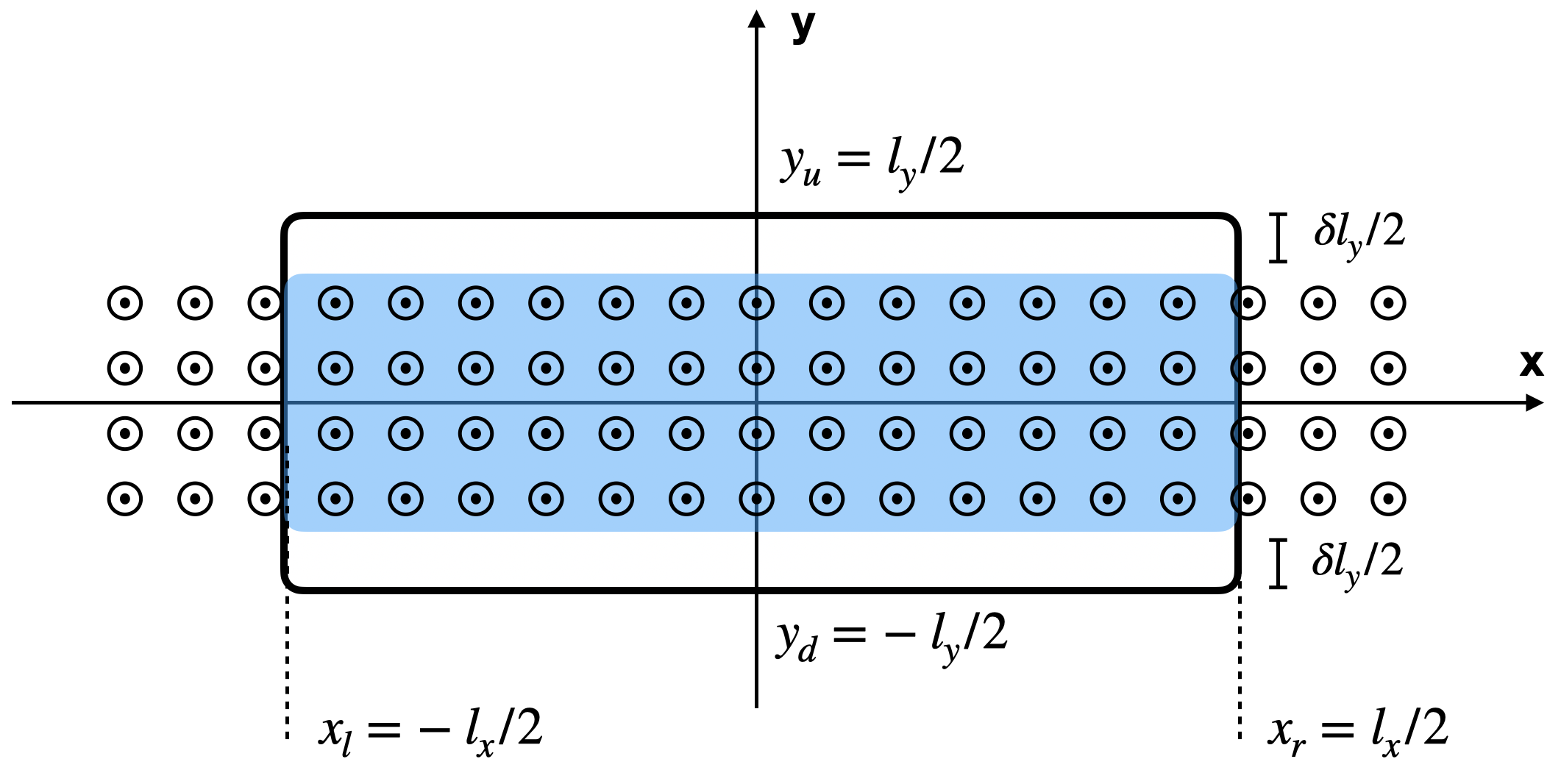}\\
\caption{\textbf{Setup configuration}. The conducting loop is represented by the thick line, and its shape approximates a rectangle of height $l_{\textrm{y}}:=y_{\textrm{d}}-y_{\textrm{u}}$ and width $l_{\textrm{x}}:=x_{\textrm{r}}-x_{\textrm{l}}$. The magnetic field $\boldsymbol{B}=B_0\hat{\mathbf{z}}$ is non vanishing within the light-blue shaded area, which leaves a gap of length $\delta l_{\text{y}}$ on both ends in the $y$-direction between the top and bottom edges of the loop, it is perpendicular to the plane and points outwards.}\label{Figure:Zero:Appendix}
\end{figure}

The wire is assumed to be infinitely rigid, which means that it has a constant proper length \cite{Lyle:2010}. We introduce the proper length between two points $x^\mu_{\textrm{i}}$ and $x^\mu_{\textrm{f}}$ in spacetime as $L:=\int_{x^\mu_{\textrm{i}}}^{x^\mu_{\textrm{f}}}\sqrt{-ds^2}$ along a trajectory $x^\mu(\lambda)$ parametrized by $\lambda$. We thus compute $L_{\textrm{x}}:=\int_{x_{\textrm{l}}}^{x_{\textrm{r}}}\sqrt{-ds^2}|_{y=z=0}$ and $L_{\textrm{y}}:=\int_{y_{\textrm{d}}}^{y_{\textrm{u}}}\sqrt{-ds^2}|_{x=z=0}$ at constant time $t$.

\subsection{Proper length of the coil: general expression}
In the case of the general perturbation, we find
\begin{align}\label{proper:lengths:general:appendix}
L_{\textrm{x}}(t)=&l_{\textrm{x}}\left(1+\epsilon\Lambda_{\textrm{x}\textrm{x}}(t)\right),\nonumber\\
L_{\textrm{y}}(t)=&l_{\textrm{y}}\left(1+\epsilon\Lambda_{\textrm{y}\textrm{y}}(t)\right),
\end{align}
where we have defined the real and time-dependent functions $\Lambda_{\textrm{x}\textrm{x}}(t):=(1/l_{\textrm{x}})\int_{x_{\textrm{l}}}^{x_{\textrm{r}}}dx\, h_{\textrm{x}\textrm{x}}(x^\rho)|_{y=z=0}$ and $\Lambda_{\textrm{y}\textrm{y}}(t):=(1/l_{\textrm{y}})\int_{y_{\textrm{d}}}^{y_{\textrm{u}}} dy\, h_{\textrm{y}\textrm{y}}(x^\rho)|_{x=z=0}$. Note that, in general, the proper lengths $L_{\textrm{x}}(t)$ and $L_{\textrm{y}}(t)$ will not coincide with the coordinate lengths $l_{\textrm{x}}$ and $l_{\textrm{y}}$. This occurs at least in the case of (asymptotically) flat spacetime.

\subsection{Proper length of the coil: radiating perturbation}
In the case of the radiating perturbation we obtain
\begin{align}\label{proper:lengths:radiative:perturbation:appendix}
L_{\textrm{x}}(t)=&l_{\textrm{x}}\left(1+\epsilon \Re\int d^3k \, C_{\textrm{x}\textrm{x}}(\bold{k})e^{-ic|\bold{k}|t}\textrm{sinc}\bigl(k_{\textrm{x}}l_{\textrm{x}}\bigr)\right),\nonumber\\
L_{\textrm{y}}(t)=&l_{\textrm{y}}\left(1+\epsilon \Re\int d^3k \, C_{\textrm{y}\textrm{y}}(\bold{k})e^{-ic|\bold{k}|t}\textrm{sinc}\bigl(k_{\textrm{y}}l_{\textrm{y}}\bigr)\right).
\end{align}

\section{ELECTRODYNAMICS IN (WEAKLY) CURVED SPACETIME}\label{electrodynamics:appendix}
Here we derive our main equations \eqref{main:relations}. We refer the reader to the literature for the introduction to the topics and tools employed for this purpose \cite{Thorne:MacDonald:1982,Cabral:Lobo:2017:v2,Cabral:Lobo:2017:v2,Brunney:Gradoni:2021}.

We start by assuming that we are working in weakly curved Schwarzschild spacetime, which well models the spacetime outside a nonrotating (or slowly rotating) spherical planet, such as the Earth. The typical dimensionless strength $r_\text{S}/$ of the local gravitational potential will be of the order of $ r_\text{S}/r\sim10^{-9}$ or less, where $r_\text{S}$ is the Schwarzschild radius for the massive planet and $r$ is the distance of the point of measurement from the centre of the Earth. This means that we can establish, to lowest order, a set of preferential observers, namely the inertial observers, that will be performing the measurements in their laboratories.  

Electrodynamics in curved spacetime can be implemented by considering the 4-vector field $A^\nu$ that defines the Faraday tensor $F_{\mu\nu}:=\nabla_\mu A_\nu-\nabla_\nu A_\mu$ and the 4-current $J^\nu$ that is conserved, i.e., that satisfies $\nabla_\mu J^\mu=0$, Here $\nabla_\nu$ is the covariant derivative. The dynamical equations for the field read 
\begin{align}\label{main:equation:appendix}
\nabla_\sigma(\nabla^\sigma A_\nu)-\nabla_\sigma(\nabla_\nu A^\sigma)=-R_{\sigma\nu}\,A^\sigma+\mu_0\,J_\nu,
\end{align}
which, since we choose to work in the Lorentz gauge $\nabla_\sigma A^\sigma=0$, reduce to
\begin{align}\label{main:equation:two:appendix}
\nabla_\sigma(\nabla^\sigma A_\nu)=-R_{\sigma\nu}\,A^\sigma+\mu_0\,J_\nu.
\end{align}
Two useful and equivalent expressions to \eqref{main:equation:appendix} are $\square A_\sigma+\frac{1}{\sqrt{-g}}\partial_\mu\left(\sqrt{-g}\,g^{\mu\nu}\right)\,\partial_\nu\,A_\sigma+g_{\sigma\mu}R^{\mu\nu}\,A_\nu=\mu_0\,J_\sigma$ and $(\sqrt{-g})^{-1}\partial_\mu F^{\mu\nu}\partial_\mu(\sqrt{-g})\,F^{\mu\nu}=\mu_0\,J^\nu$.

We also assume that the only ``free charges'' are confined in the wire. A conducting wire, in general, has free electrons and free positive charges (atoms that lose an electron to the conductive band) that all together compensate on average \cite{Griffiths:2017}. This means that we can safely assume that $J^\mu=0$. This is consistent with considerations made in this direction that are found in the literature \cite{Fortini:Gualdi:1991}.

This assumption further allows us to write
\begin{align}
\nabla_\sigma(\nabla^\sigma A_\nu)=-R_{\sigma\nu}\,A^\sigma.
\end{align}
We have obtained a simplified homogeneous field equation \eqref{main:equation:two:appendix} in vacuum and we know that the connection is torsionless in the case of our spacetime. In the literature (below Eq.(3.26) in \cite{Cabral:Lobo:2017:v2} and Eq.(4.3) in \cite{Thorne:MacDonald:1982}), it has been shown that this implies that Faraday's and Lenz's law read
\begin{align}\label{main:relations:appendix}
\Phi_\textrm{M}=&\int_{\Sigma} d\Sigma^j\, B_j\nonumber\\
\dot\Phi_\textrm{M}=&-\oint_{\partial_{\Sigma}} dx^k\, j^\Phi_k,
\end{align}
where $\Sigma$ is the two-surface of integration with element $d\Sigma^j$, $\partial_\Sigma$ is the boundary of $\Sigma$ and $dx^k$ is the element along $\partial_\Sigma$. Notice that, as argued in \cite{Thorne:MacDonald:1982}, the integrals  are performed along the ``proper dimensions'' of the surface or the boundary.

\section{DERIVATION OF THE MAIN EQUATIONS}

We ask ourselves how does the Faraday induction law in \eqref{main:relations:appendix} read in this case. We do not have to choose a particular surface through which the flux is defined, and our work can be immediately adapted to different surfaces. We choose here a flat surface to illustrate the principle. In this case, we assume a flat surface that lies in the $xy$ plane. We note that the oriented volume element $d\Sigma^j=\sqrt{g_2}\,dx\,dy\,\hat{\boldsymbol{z}}$ can have a dependence on the metric. In principle, the reduced metric determinant $g_2$ can have first-order contributions, therefore the Faraday induction law will take the general form
\begin{align}\label{magnetic:flux:appendix}
\Phi_\textrm{M}(t):=\Phi_\textrm{M}^{(0)}(t)+\epsilon\int_{\Sigma^{(1)}} \,dx\,dy\, B_\text{z}^{(0)}+\epsilon\int_{\Sigma^{(0)}}dx\,dy\, g_2^{(1)}(t,x,y)\,B_\text{z}^{(0)}.
\end{align}
Here, the first contribution is just the lowest order flat spacetime contribution, while the second and third terms includes the flat spacetime flux through the unperturbed surface $\Sigma^{(0)}$, while $\Sigma^{(1)}$ is the deformation induced on the surface due to gravity. 

\subsection{Derivation of the main result for a general linearized metric}\label{derivation:main:result:appendix}
We now proceed to compute the current induced by the generic change in time of the gravitational field using the main equations \eqref{main:relations:appendix}. We need to define precisely the surface $\Sigma$ of interest. We have assumed that the wire of conducting material is infinitely rigid, therefore, the integration region $\Sigma$ is defined by $-L_{\textrm{x}}/2\leq x\leq L_{\textrm{x}}/2$ and $-L_{\textrm{y}}/2\leq y\leq L_{\textrm{y}}/2$. We also need the expression $l_{\textrm{y}}(t)=L_{\textrm{y}}\left(1-\epsilon\Lambda_{\textrm{y}\textrm{y}}(t)\right)$ obtained from those found in \eqref{proper:lengths:general:appendix}.
We compute $\Phi_\textrm{M}(t)$ explicitly to first order as shown below
\begin{align}
\Phi_\textrm{M}(t)=&\int_{-L_{\textrm{x}}/2}^{L_{\textrm{x}}/2}dx \int_{-L_{\textrm{y}}/2}^{L_{\textrm{y}}/2}dy\,\left(1+\epsilon g_2^{(1)}(t,x,y)\right)B(y)\nonumber\\
=&B_0 L_{\textrm{x}} \int_{-L_{\textrm{y}}/2}^{L_{\textrm{y}}/2} dy\,(\vartheta(y+(l_{\textrm{y}}/2-\delta l_{\text{y}}/2))-\vartheta(y-(l_{\textrm{y}}/2-\delta l_{\text{y}}/2)))\nonumber\\
&+\epsilon\,B_0\int_{-l_{\textrm{x}}/2}^{l_{\textrm{x}}/2}dx\int_{-l_{\textrm{y}}/2}^{l_{\textrm{y}}/2} dy\,(\vartheta(y+(l_{\textrm{y}}/2-\delta l_{\text{y}}/2))-\vartheta(y-(l_{\textrm{y}}/2-\delta l_{\text{y}}/2))) g_2^{(1)}(t,x,y)\nonumber\\
=&B_0 L_{\textrm{x}} \int_{-L_{\textrm{y}}/2}^{L_{\textrm{y}}/2} dy\,(\vartheta(y+((1-\epsilon \Lambda_{\textrm{y}\textrm{y}}(t))L_{\textrm{y}}/2-\delta l_{\text{y}}/2))-\vartheta(y-((1-\Lambda_{\textrm{y}\textrm{y}}(t))L_{\textrm{y}}/2-\delta l_{\text{y}}/2)))\nonumber\\
&+\epsilon\,B_0\frac{l_{\textrm{x}}l_{\textrm{y}}}{4}\int_{-1}^1d\tilde{x}\int_{-1}^1 d\tilde{y}\,(\vartheta(\tilde{y}+(1-\epsilon_{\text{y}}))-\vartheta(\tilde{y}-(1-\epsilon_{\text{y}}))) \int d^2k\,\tilde{g}_2^{(1)}(t,k_{\textrm{x}},k_{\textrm{y}})\,e^{i\frac{l_{\textrm{x}}}{2}k_{\textrm{x}}\tilde{x}}e^{i\frac{l_{\textrm{y}}}{2}k_{\textrm{y}}\tilde{y}}\nonumber\\
=&B_0 \frac{L_{\textrm{x}}L_{\textrm{y}}}{2} \int_{-1}^1 d\tilde{y}\,\vartheta(\tilde{y}+(1-\epsilon \Lambda_{\textrm{y}\textrm{y}}(t))(1-\epsilon_{\text{y}}))-B_0 \frac{L_{\textrm{x}}L_{\textrm{y}}}{2} \int_{-1}^1 d\tilde{y}\,\vartheta(\tilde{y}-(1-\Lambda_{\textrm{y}\textrm{y}}(t))(1-\epsilon_{\text{y}}))\nonumber\\
&+\epsilon\,B_0\frac{l_{\textrm{x}}l_{\textrm{y}}}{2}\int_{-1}^1 d\tilde{y}\,\vartheta(\tilde{y}+(1-\epsilon_{\text{y}})) \int d^2k\,\tilde{g}_2^{(1)}(t,k_{\textrm{x}},k_{\textrm{y}})\,\textrm{sinc}\left(\frac{l_{\textrm{x}}}{2}k_{\textrm{x}}\right)\,e^{i\frac{l_{\textrm{y}}}{2}k_{\textrm{y}}\tilde{y}}\nonumber\\
&-\epsilon\,B_0\frac{l_{\textrm{x}}l_{\textrm{y}}}{2}\int_{-1}^1 d\tilde{y}\,\vartheta(\tilde{y}-(1-\epsilon_{\text{y}})) \int d^2k\,\tilde{g}_2^{(1)}(t,k_{\textrm{x}},k_{\textrm{y}})\,\textrm{sinc}\left(\frac{l_{\textrm{x}}}{2}k_{\textrm{x}}\right)\,e^{i\frac{l_{\textrm{y}}}{2}k_{\textrm{y}}\tilde{y}}\nonumber\\
=&B_0 \frac{L_{\textrm{x}}L_{\textrm{y}}}{2} \int_{-1}^1 d\tilde{y}\,\vartheta(\tilde{y}+(1-\epsilon \Lambda_{\textrm{y}\textrm{y}}(t))(1-\epsilon_{\text{y}}))-B_0 \frac{L_{\textrm{x}}L_{\textrm{y}}}{2} \int_{-1}^1 d\tilde{y}\,\vartheta(\tilde{y}-(1-\Lambda_{\textrm{y}\textrm{y}}(t))(1-\epsilon_{\text{y}}))\nonumber\\
&+\epsilon\,B_0\frac{l_{\textrm{x}}l_{\textrm{y}}}{2}\int_{-1}^1 d\tilde{y}\,\vartheta(\tilde{y}+(1-\epsilon_{\text{y}})) \int d^2\tilde{k}\,\tilde{g}_2^{(1)\prime}(t,\tilde{k}_{\textrm{x}},\tilde{k}_{\textrm{y}})\,\textrm{sinc}\bigl(\tilde{k}_{\textrm{x}}\bigr)\,e^{i\tilde{k}_\text{y}\tilde{y}}\nonumber\\
&-\epsilon\,B_0\frac{l_{\textrm{x}}l_{\textrm{y}}}{2}\int_{-1}^1 d\tilde{y}\,\vartheta(\tilde{y}-(1-\epsilon_{\text{y}})) \int d^2\tilde{k}\,\tilde{g}_2^{(1)\prime}(t,\tilde{k}_{\textrm{x}},\tilde{k}_{\textrm{y}})\,\textrm{sinc}\bigl(\tilde{k}_{\textrm{x}}\bigr)\,e^{i\tilde{k}_\text{y}\tilde{y}}\nonumber\\.
\end{align}
We have introduced $\epsilon_{\textrm{y}}:=\delta L_{\text{y}}/L_{\textrm{y}}$ and the Fourier transform $\tilde{g}_2^{(1)}(t,x,y) :=\int d^k \tilde{g}_2^{(1)}(t,k_{\textrm{x}},k_{\textrm{y}}) e^{i k_{\textrm{x}}x}e^{i k_{\textrm{y}}y}$. Here $\delta l_{\text{y}}=(1-\epsilon \Lambda_{\textrm{y}\textrm{y}}(t))\delta L_{\text{y}}$. To obtain the last line, we have used the fact that $\vartheta(\alpha x)=\vartheta(x)$ if $\alpha>0$.
In the above, we can exchange $L_{\textrm{x}}$ and $L_{\textrm{y}}$ with $l_{\textrm{x}}$ and $l_{\textrm{y}}$ when the whole expression is multiplied by $\epsilon$ by paying the price of an error to second order in $\epsilon$. We will do so below as well when necessary. 
We also notice that  $\epsilon_{\textrm{y}}=\delta L_{\text{y}}/L_{\textrm{y}}=\delta l_{\text{y}}/l_{\textrm{y}}$.

We have introduced $\tilde{g}_2^{(1)\prime}(t,\tilde{k}_{\textrm{x}},\tilde{k}_{\textrm{y}}):=\frac{4}{l_{\textrm{x}}l_{\textrm{y}}}\tilde{g}_2^{(1)}(t,\frac{2}{l_{\textrm{x}}}\tilde{k}_{\textrm{x}},\frac{2}{l_{\textrm{y}}}\tilde{k}_{\textrm{y}})$ for convenience of presentation and $\tilde{k}_{\textrm{x}},\tilde{k}_{\textrm{y}}$ are dimensionless variables.
We continue and obtain
{\small
\begin{align}
\Phi_\textrm{M}(t)=&B_0 \frac{L_{\textrm{x}}L_{\textrm{y}}}{2}\left\{ \int_0^2 d\tilde{y}\,\vartheta(\tilde{y}-\epsilon \Lambda_{\textrm{y}\textrm{y}}(t)-\epsilon_{\text{y}})- \int_{-2}^0 d\tilde{y}\,\vartheta(\tilde{y}+\epsilon\Lambda_{\textrm{y}\textrm{y}}(t)+\epsilon_{\text{y}})\right.\nonumber\\
+&\left.\epsilon\,\int_0^2 d\tilde{y}\,\vartheta(\tilde{y}-\epsilon_{\text{y}}) \int d^2\tilde{k}\,\tilde{g}_2^{(1)\prime}(t,\tilde{k}_{\textrm{x}},\tilde{k}_{\textrm{y}})\,\textrm{sinc}\bigl(\tilde{k}_{\textrm{x}}\bigr)\, e^{-i\tilde{k}_{\textrm{y}}}e^{i\tilde{k}_{\textrm{y}}\tilde{y}}-\epsilon\,\int_{-2}^0 d\tilde{y}\,\vartheta(\tilde{y}+\epsilon_{\text{y}}) \int d^2\tilde{k}\,\tilde{g}_2^{(1)\prime}(t,\tilde{k}_{\textrm{x}},\tilde{k}_{\textrm{y}})\,\textrm{sinc}\bigl(\tilde{k}_{\textrm{x}}\bigr)\,e^{i\tilde{k}_{\textrm{y}}} e^{i\tilde{k}_{\textrm{y}}\tilde{y}}\right\}\nonumber\\
=&B_0 \frac{L_{\textrm{x}}L_{\textrm{y}}}{2}\left\{(2-(\epsilon\Lambda_{\textrm{y}\textrm{y}}(t)+\epsilon_{\text{y}})\vartheta(\epsilon\Lambda_{\textrm{y}\textrm{y}}(t)+\epsilon_{\text{y}}))- (0+(\epsilon\Lambda_{\textrm{y}\textrm{y}}(t)+\epsilon_{\text{y}})\vartheta(\epsilon\Lambda_{\textrm{y}\textrm{y}}(t)+\epsilon_{\text{y}}))\right.\nonumber\\
+&\left.\epsilon\,\int_{\epsilon_{\text{y}}\vartheta(\epsilon_{\text{y}})}^2 d\tilde{y}\,\int d^2\tilde{k}\,\tilde{g}_2^{(1)\prime}(t,\tilde{k}_{\textrm{x}},\tilde{k}_{\textrm{y}})\,\textrm{sinc}\bigl(\tilde{k}_{\textrm{x}}\bigr)\, e^{-i\tilde{k}_{\textrm{y}}}e^{i\tilde{k}_{\textrm{y}}\tilde{y}}-\epsilon\,\int_{-\epsilon_{\text{y}}\vartheta(\epsilon_{\text{y}})}^0 d\tilde{y}\,\int d^2\tilde{k}\,\tilde{g}_2^{(1)\prime}(t,\tilde{k}_{\textrm{x}},\tilde{k}_{\textrm{y}})\,\textrm{sinc}\bigl(\tilde{k}_{\textrm{x}}\bigr)\, e^{i\tilde{k}_{\textrm{y}}}e^{i\tilde{k}_{\textrm{y}}\tilde{y}}\right\}.
\end{align}
}
Here, we have noted that the lower bound of the second integral in the first line becomes $-(\Lambda_{\textrm{y}\textrm{y}}(t)+\epsilon_{\text{y}})\vartheta(\Lambda_{\textrm{y}\textrm{y}}(t)+\epsilon_{\text{y}})$ instead of $-2$. In fact, if $\Lambda_{\textrm{y}\textrm{y}}(t)+\epsilon_{\text{y}}<0$ it then follows that $\vartheta(\tilde{y}+\Lambda_{\textrm{y}\textrm{y}}(t)+\epsilon_{\text{y}})=0$ since $\tilde{y}\leq0$. The same reasoning applies to the integrals in the second and third line.
Finally, we clean up the previous expression and proceed to obtain
{\small
\begin{align}\label{final:expression:flux:general:appendix}
\Phi_\textrm{M}(t)=\Phi_\textrm{M}(0)\left[1-(\epsilon\Lambda_{\textrm{y}\textrm{y}}(t)+\epsilon_{\text{y}})\vartheta(\epsilon\Lambda_{\textrm{y}\textrm{y}}(t)+\epsilon_{\text{y}})+\epsilon (1-\epsilon_{\text{y}}\vartheta(\epsilon_{\text{y}})) \int d^2\tilde{k}\,\tilde{g}_2^{(1)\prime}(t,\tilde{k}_{\textrm{x}},\tilde{k}_{\textrm{y}}) \textrm{sinc}(\tilde{k}_{\textrm{x}}) \textrm{sinc}\left(\tilde{k}_{\textrm{y}} (1-\epsilon_{\text{y}}\vartheta(\epsilon_{\text{y}}))\right)\right],
\end{align}
}
where we have introduced the flux constant $\Phi_\textrm{M}(0)=A B_0=B_0 L_{\textrm{x}}L_{\textrm{y}}$.

\subsection{Consequences of the presence of the gap}
We here wish to employ the expression \eqref{final:expression:flux:general:appendix} and analyze the influence of the gap $\epsilon_\text{y}$ on the flux.

\textbf{Case (I): $\epsilon_\text{y}\geq0$}. In this case, we note that $1-\epsilon_{\text{y}}\vartheta(\epsilon_{\text{y}})=1-\epsilon_{\text{y}}\geq0$ since we always have $\epsilon_\text{y}\leq1$. The interesting part is the term proportional to $Z(t):=(\epsilon\Lambda_{\textrm{y}\textrm{y}}(t)+\epsilon_{\text{y}})\vartheta(\epsilon\Lambda_{\textrm{y}\textrm{y}}(t)+\epsilon_{\text{y}})$. Clearly, when $\Lambda_{\textrm{y}\textrm{y}}(t)\geq-\epsilon_{\text{y}}/\epsilon$ we have that $\vartheta(\epsilon\Lambda_{\textrm{y}\textrm{y}}(t)+\epsilon_{\text{y}})=1$ and thus $Z(t)=\epsilon\Lambda_{\textrm{y}\textrm{y}}(t)+\epsilon_{\text{y}}$. However, if there are two times $t_1$ and $t_2$ between which $\Lambda_{\textrm{y}\textrm{y}}(t)<-\epsilon_{\text{y}}/\epsilon$, it then occurs that $Z(t)=0$ for $t_1\leq t\leq t_2$.

\textbf{Case (II): $\epsilon_\text{y}<0$}. In this case, we note that $1+|\epsilon_{\text{y}}|\vartheta(-|\epsilon_{\text{y}}|)=0$. The only remaining part is the term proportional to $Z(t):=(\epsilon\Lambda_{\textrm{y}\textrm{y}}(t)-|\epsilon_{\text{y}}|)\vartheta(\epsilon\Lambda_{\textrm{y}\textrm{y}}(t)-|\epsilon_{\text{y}}|)$. Clearly, when $\Lambda_{\textrm{y}\textrm{y}}(t)\geq|\epsilon_{\text{y}}|/\epsilon$ we have that $\vartheta(\epsilon\Lambda_{\textrm{y}\textrm{y}}(t)-|\epsilon_{\text{y}}|)=1$ and thus $Z(t)=\epsilon\Lambda_{\textrm{y}\textrm{y}}(t)-|\epsilon_{\text{y}}|$. Therefore, if there are two times $t_1$ and $t_2$ between which $\Lambda_{\textrm{y}\textrm{y}}(t)<|\epsilon_{\text{y}}|/\epsilon$, it then occurs that $Z(t)=0$ for $t_1\leq t\leq t_2$.

\subsection{Derivation of the main result for a radiating dynamical perturbation}
We apply the general results above to the case of radiating dynamical perturbations. We will need to inverted expression $l_{\textrm{y}}(t)=L_{\textrm{y}}\left(1-\epsilon \Re\int d^3k \, C_{\textrm{y}\textrm{y}}(\bold{k})e^{-ic|\bold{k}|t}\phi_{\textrm{y}}\right)$ and the expression $g_2^{(1)}(t,x,y)|_{z=0}=h_\text{xx}|_{z=0}+h_\text{yy}|_{z=0}=-h_\text{zz}|_{z=0}$, since Tr$(h_{\mu\nu})=0$.
We find
\begin{align}
\Phi_\textrm{M}(t)=&\int_{-L_{\textrm{x}}/2}^{L_{\textrm{x}}/2}dx \int_{-L_{\textrm{y}}/2}^{L_{\textrm{y}}/2}dy\,\left(1-\epsilon\Re\int d^3k\,C_{\textrm{z}\textrm{z}}(\bold{k})\,e^{k_\sigma x^\sigma}|_{z=0}\right)B(y)\nonumber\\
=&B_0 L_{\textrm{x}} \int_{-L_{\textrm{y}}/2}^{L_{\textrm{y}}/2} dy\,(\vartheta(y+(l_{\textrm{y}}/2-\delta l_{\text{y}}/2))-\vartheta(y-(l_{\textrm{y}}/2-\delta l_{\text{y}}/2)))\nonumber\\
&-\epsilon\,B_0\int_{-l_{\textrm{x}}/2}^{l_{\textrm{x}}/2}dx\int_{-l_{\textrm{y}}/2}^{l_{\textrm{y}}/2} dy\,(\vartheta(y+(l_{\textrm{y}}/2-\delta l_{\text{y}}/2))-\vartheta(y-(l_{\textrm{y}}/2-\delta l_{\text{y}}/2))) \Re\int d^3k\,C_{\textrm{z}\textrm{z}}(\bold{k})\,e^{k_\sigma x^\sigma}|_{z=0}\nonumber\\
=&B_0 L_{\textrm{x}} \int_{-L_{\textrm{y}}/2}^{L_{\textrm{y}}/2} dy\,(\vartheta(y+(1-\epsilon \Lambda_{\textrm{y}\textrm{y}}(t))(L_{\textrm{y}}/2-\delta L_{\text{y}}/2))-\vartheta(y-(1-\Lambda_{\textrm{y}\textrm{y}}(t))(L_{\textrm{y}}/2-\delta L_{\text{y}}/2)))\nonumber\\
&-\epsilon\frac{\Phi_\textrm{M}(0)}{4}\int_{-1}^1d\tilde{x} d\tilde{y}\,(\vartheta(\tilde{y}+(1-\epsilon_{\text{y}}))-\vartheta(\tilde{y}-(1-\epsilon_{\text{y}}))) \Re\int d^3k\,C_{\textrm{z}\textrm{z}}(\bold{k})\,e^{-ic|\bold{k}|t}e^{i\frac{l_{\textrm{x}}}{2}k_{\textrm{x}}\tilde{x}}e^{i\frac{l_{\textrm{y}}}{2}k_{\textrm{y}}\tilde{y}}\nonumber\\
=& \frac{\Phi_\textrm{M}(0)}{2} \int_{-1}^1 d\tilde{y}\,\vartheta(\tilde{y}+(1-\epsilon \Lambda_{\textrm{y}\textrm{y}}(t))(1-\epsilon_{\text{y}}))- \frac{\Phi_\textrm{M}(0)}{2} \int_{-1}^1 d\tilde{y}\,\vartheta(\tilde{y}-(1-\Lambda_{\textrm{y}\textrm{y}}(t))(1-\epsilon_{\text{y}}))\nonumber\\
&-\epsilon\frac{\Phi_\textrm{M}(0)}{4}\int_{-1}^1d\tilde{x} d\tilde{y}\,(\vartheta(\tilde{y}+(1-\epsilon_{\text{y}}))-\vartheta(\tilde{y}-(1-\epsilon_{\text{y}}))) \Re\int d^3k\,C_{\textrm{z}\textrm{z}}(\bold{k})\,e^{-ic|\bold{k}|t}e^{i\frac{l_{\textrm{x}}}{2}k_{\textrm{x}}\tilde{x}}e^{i\frac{l_{\textrm{y}}}{2}k_{\textrm{y}}\tilde{y}}\nonumber\\
=&\frac{\Phi_\textrm{M}(0)}{2} \int_{-1}^1 d\tilde{y}\,\vartheta(\tilde{y}+(1-\epsilon \Lambda_{\textrm{y}\textrm{y}}(t)-\epsilon_{\text{y}}))- \frac{\Phi_\textrm{M}(0)}{2} \int_{-1}^1 d\tilde{y}\,\vartheta(\tilde{y}-(1-\Lambda_{\textrm{y}\textrm{y}}(t)-\epsilon_{\text{y}}))\nonumber\\
&-\epsilon\frac{\Phi_\textrm{M}(0)}{2}\int_{-1}^1 d\tilde{y}\,\vartheta(\tilde{y}+(1-\epsilon_{\text{y}})) \Re\int d^3k\,C_{\textrm{z}\textrm{z}}(\bold{k})\,\textrm{sinc}\left(\frac{l_{\textrm{x}}}{2}k_{\textrm{x}}\right)\,e^{-ic|\bold{k}|t}e^{i\frac{l_{\textrm{y}}}{2}k_{\textrm{y}}\tilde{y}}\nonumber\\
&+\epsilon \frac{\Phi_\textrm{M}(0)}{2}\int_{-1}^1 d\tilde{y}\,\vartheta(\tilde{y}-(1-\epsilon_{\text{y}})) \Re\int d^3k\,C_{\textrm{z}\textrm{z}}(\bold{k})\,\textrm{sinc}\left(\frac{l_{\textrm{x}}}{2}k_{\textrm{x}}\right)\,e^{-ic|\bold{k}|t}e^{i\frac{l_{\textrm{y}}}{2}k_{\textrm{y}}\tilde{y}}
\end{align}
Here we have used the fact that $\epsilon l_{\textrm{x}}l_{\textrm{y}}=\epsilon L_{\textrm{x}}L_{\textrm{y}}$ to first order and we have ignored terms of the form $\epsilon\epsilon_{\text{y}}$.

As done above, we continue by shifting the limits of integration. We have
{\small
\begin{align}
\Phi_\textrm{M}(t)=&\frac{\Phi_\textrm{M}(0)}{2} \int_0^2 d\tilde{y}\,\vartheta(\tilde{y}-\epsilon \Lambda_{\textrm{y}\textrm{y}}(t)-\epsilon_{\text{y}})- \frac{\Phi_\textrm{M}(0)}{2} \int_{-2}^0 d\tilde{y}\,\vartheta(\tilde{y}+\epsilon\Lambda_{\textrm{y}\textrm{y}}(t)+\epsilon_{\text{y}})\nonumber\\
&-\epsilon\,\frac{\Phi_\textrm{M}(0)}{2}\int_0^2 d\tilde{y}\,\vartheta(\tilde{y}-\epsilon_{\text{y}}) \Re\int d^3k\,C_{\textrm{z}\textrm{z}}(\bold{k})\,\textrm{sinc}\left(\frac{l_{\textrm{x}}}{2}k_{\textrm{x}}\right)\,e^{-ic|\bold{k}|t} e^{-i\frac{l_{\textrm{y}}}{2}k_{\textrm{y}}}e^{i\frac{l_{\textrm{y}}}{2}k_{\textrm{y}}\tilde{y}}\nonumber\\
&+\epsilon\,\frac{\Phi_\textrm{M}(0)}{2}\int_{-2}^0 d\tilde{y}\,\vartheta(\tilde{y}+\epsilon_{\text{y}}) \Re\int d^3k\,C_{\textrm{z}\textrm{z}}(\bold{k})\,\textrm{sinc}\left(\frac{l_{\textrm{x}}}{2}k_{\textrm{x}}\right)\,e^{-ic|\bold{k}|t} e^{i\frac{l_{\textrm{y}}}{2}k_{\textrm{y}}}e^{i\frac{l_{\textrm{y}}}{2}k_{\textrm{y}}\tilde{y}}\nonumber\\
=&B_0\frac{L_{\textrm{x}}L_{\textrm{y}}}{2}(2-(\epsilon\Lambda_{\textrm{y}\textrm{y}}(t)+\epsilon_{\text{y}})\vartheta(\epsilon\Lambda_{\textrm{y}\textrm{y}}(t)+\epsilon_{\text{y}}))-B_0 \frac{L_{\textrm{x}}L_{\textrm{y}}}{2} (0+(\epsilon\Lambda_{\textrm{y}\textrm{y}}(t)+\epsilon_{\text{y}})\vartheta(\epsilon\Lambda_{\textrm{y}\textrm{y}}(t)+\epsilon_{\text{y}}))\nonumber\\
&-\epsilon\,B_0\frac{L_{\textrm{x}}L_{\textrm{y}}}{2}\,\int_{\epsilon_{\text{y}}\vartheta(\epsilon_{\text{y}})}^2 d\tilde{y}\,\Re\int d^3k\,C_{\textrm{z}\textrm{z}}(\bold{k})\,\textrm{sinc}\left(\frac{l_{\textrm{x}}}{2}k_{\textrm{x}}\right)\,e^{-ic|\bold{k}|t} e^{-i\frac{l_{\textrm{y}}}{2}k_{\textrm{y}}}e^{i\frac{l_{\textrm{y}}}{2}k_{\textrm{y}}\tilde{y}}\nonumber\\
&+\epsilon\,B_0\frac{L_{\textrm{x}}L_{\textrm{y}}}{2}\,\int_{-\epsilon_{\text{y}}\vartheta(\epsilon_{\text{y}})}^0 d\tilde{y}\,\Re\int d^3k\,C_{\textrm{z}\textrm{z}}(\bold{k})\,\textrm{sinc}\left(\frac{l_{\textrm{x}}}{2}k_{\textrm{x}}\right)\,e^{-ic|\bold{k}|t} e^{i\frac{l_{\textrm{y}}}{2}k_{\textrm{y}}}e^{i\frac{l_{\textrm{y}}}{2}k_{\textrm{y}}\tilde{y}}.\nonumber\\
=&\frac{\Phi_\textrm{M}(0)}{2}(2-(\epsilon\Lambda_{\textrm{y}\textrm{y}}(t)+\epsilon_{\text{y}})\vartheta(\epsilon\Lambda_{\textrm{y}\textrm{y}}(t)+\epsilon_{\text{y}}))-\frac{\Phi_\textrm{M}(0)}{2} (0+(\epsilon\Lambda_{\textrm{y}\textrm{y}}(t)+\epsilon_{\text{y}})\vartheta(\epsilon\Lambda_{\textrm{y}\textrm{y}}(t)+\epsilon_{\text{y}}))\nonumber\\
&-\epsilon\,\frac{\Phi_\textrm{M}(0)}{2}\,\int_{\epsilon_{\text{y}}\vartheta(\epsilon_{\text{y}})}^2 d\tilde{y}\,\Re\int d^3\tilde{k}\,\tilde{C}'_{\textrm{z}\textrm{z}}(\tilde{\bold{k}})\,\textrm{sinc}\bigl(\tilde{k}_{\textrm{x}}\bigr)\,e^{-i\tilde{\omega}_{\tilde{\bold{k}}}t} e^{-i\tilde{k}_{\textrm{y}}}e^{i\tilde{k}_{\textrm{y}}\tilde{y}}\nonumber\\
&+\epsilon\,\frac{\Phi_\textrm{M}(0)}{2}\,\int_{-\epsilon_{\text{y}}\vartheta(\epsilon_{\text{y}})}^0 d\tilde{y}\,\Re\int d^3\tilde{k}\,\tilde{C}'_{\textrm{z}\textrm{z}}(\tilde{\bold{k}})\,\textrm{sinc}\bigl(\tilde{k}_{\textrm{x}}\bigr)\,e^{-i\tilde{\omega}_{\tilde{\bold{k}}}t} e^{i\tilde{k}_{\textrm{y}}}e^{i\tilde{k}_{\textrm{y}}\tilde{y}}.
\end{align}
}
Here, we have defined $\tilde{C}'_{\textrm{z}\textrm{z}}(\tilde{\bold{k}}):=\frac{8}{l_{\textrm{x}}l_{\textrm{y}}\sqrt{l_{\textrm{x}}l_{\textrm{y}}}}\tilde{C}_{\textrm{z}\textrm{z}}\bigl(\frac{2}{l_{\textrm{x}}}\tilde{k}_\text{x},\frac{2}{l_{\textrm{y}}}\tilde{k}_\text{y},\frac{2}{\sqrt{l_{\textrm{x}}l_{\textrm{y}}}}\tilde{k}_\text{z}\bigr)$ and $\tilde{\omega}_{\tilde{\bold{k}}}:=\frac{2c}{\sqrt{l_{\textrm{x}}l_{\textrm{y}}}}\sqrt{\frac{l_{\textrm{y}}}{l_{\textrm{x}}}\tilde{k}_\text{x}^2+\frac{l_{\textrm{x}}}{l_{\textrm{y}}}\tilde{k}_\text{y}+\tilde{k}_\text{z}^2}$ for convenience. We have also noted that the lower bound of the second integral in the first line becomes $-(\Lambda_{\textrm{y}\textrm{y}}(t)+\epsilon_{\text{y}})\vartheta(\Lambda_{\textrm{y}\textrm{y}}(t)+\epsilon_{\text{y}})$ instead of 2. In fact, if $\Lambda_{\textrm{y}\textrm{y}}(t)+\epsilon_{\text{y}}<0$ it then follows that $\vartheta(\tilde{y}+\Lambda_{\textrm{y}\textrm{y}}(t)+\epsilon_{\text{y}})=0$ since $\tilde{y}\leq0$. The same reasoning applies to the integrals in the second and third line. Finally, we have
{\small
\begin{align}\label{main:equation:electromagnetism:final:appendix}
\Phi_\textrm{M}(t)=&\Phi_\textrm{M}(0)\left[1-(\epsilon\Lambda_{\textrm{y}\textrm{y}}(t)+\epsilon_{\text{y}})\vartheta(\epsilon\Lambda_{\textrm{y}\textrm{y}}(t)+\epsilon_{\text{y}})-\epsilon(1-\epsilon_{\text{y}}\vartheta(\epsilon_{\text{y}}))\,\Re\int d^3\tilde{k}\,\tilde{C}'_{\textrm{z}\textrm{z}}(\tilde{\bold{k}})\,\textrm{sinc}\bigl(\tilde{k}_{\textrm{x}}\bigr)\text{sinc}\bigl((1-\epsilon_{\text{y}}\vartheta(\epsilon_{\text{y}}))\tilde{k}_\text{y}\bigr)e^{-i\tilde{\omega}_{\tilde{\bold{k}}}t}\right].
\end{align}
}
Note that if $\epsilon_{\textrm{y}}\leq0$, we would have $\vartheta(\epsilon_{\text{y}})=0$ for all times, while if $\epsilon\Lambda_{\textrm{y}\textrm{y}}(t)+\epsilon_{\text{y}}\leq0$ then we would have $\vartheta(\epsilon\Lambda_{\textrm{y}\textrm{y}}(t)+\epsilon_{\text{y}})=0$. Finally, note that the expression \eqref{main:equation:electromagnetism:final:appendix} is the equivalent of \eqref{final:expression:flux:general:appendix} specialized to the present case.

%
%

\end{document}